\documentclass[12pt,preprint]{aastex}
\usepackage{graphicx}

\def\Mdot{\hbox{$\dot {M}$}}

\def\Rsun{\hbox{\it R$_\odot$}}
\def\Zsun{\hbox{\it Z$_\odot$}}
\def\Rstar{\hbox{\it R$_*$}}
\def\Lsun{\hbox{\it L$_\odot$}}
\def\Lstar{\hbox{\it L$_*$}}
\def\Msun{\hbox{\it M$_\odot$}}

\def\Msunyr{\hbox{\it M$_\odot\,$yr$^{-1}$}}
\def\Myr{\hbox{\it Myr}}

\def\Teff{\hbox{\it T$_{\rm eff}$}}
\def\Vinf{\hbox{$v_\infty$}}
\def\kms{\hbox{km$\,$s$^{-1}$}}

\def\simgr{\mathrel{\hbox{\rlap{\hbox{\lower4pt\hbox{$\sim$}}}\hbox{$>$}}}}

\def\fnk{\hbox{\it F$_{\rm F205W}$}}
\def\fnh{\hbox{\it F$_{\rm F160W}$}}
\def\fnj{\hbox{\it F$_{\rm F110W}$}}
\def\Fnp{\hbox{\it F$_{\rm F187N}$}}
\def\Fnc{\hbox{\it F$_{\rm F190N}$}}

\def\km/s{km~s$^{-1}$}

\def\Vinf{\hbox{$V_\infty$}}
\def\HeI{He\,{\sc i}}
\def\HeII{He\,{\sc ii}}

\def\CIII{C\,{\sc iii}}
\def\CIV{C\,{\sc iv}}

\def\NIII{N\,{\sc iii}}

\def\MgII{Mg\,{\sc ii}}
\def\FeII{Fe\,{\sc ii}}

\def\SiII{Si\,{\sc ii}}
\def\Mdot{\.{M}}
\def\FMM362{FMM362}

\shorttitle{Metallicity in the Arches cluster}
\shortauthors{Najarro et al.}

\begin{document}
\title{Metallicity in the Galactic Center: \\ The Arches cluster}

\author{
Francisco Najarro\altaffilmark{1}, Donald F. Figer\altaffilmark{2}, 
D. John Hillier\altaffilmark{3}, Rolf P. Kudritzki\altaffilmark{4}}

\email{najarro@damir.iem.csic.es}

\altaffiltext{1}{Instituto de Estructura de la Materia, CSIC, Serrano 121, 29006 Madrid, Spain }
\altaffiltext{2}{Space Telescope Science Institute, 3700 San Martin Drive, Baltimore, MD 21218}
\altaffiltext{3}{Department of Physics and Astronomy, University of Pittsburgh, 3941 O'Hara Street, Pittsburgh, PA 15260}
\altaffiltext{4}{Institute for Astronomy, University of Hawaii, 2680 Woodlawn Drive, Honolulu, HI 96822}

\begin{abstract}
We present a quantitative spectral analysis of five very massive stars
in the Arches cluster, located 
near the Galactic center, to determine stellar parameters,
stellar wind properties and, most importantly, metallicity content. The
analysis uses a new technique, presented here for the first time,
and
uses line-blanketed NLTE wind/atmosphere models fit to high-resolution
near-infrared spectra of late-type nitrogen-rich Wolf-Rayet stars
and
OfI$^+$ stars in the cluster. It relies on the fact that massive stars
reach a maximum nitrogen abundance that is related to initial metallicity
when they are in the WNL phase.
We determine the present-day
nitrogen abundance of the WNL stars in the Arches cluster to be 1.6\%
(mass fraction) and constrain the stellar metallicity in the cluster to be
solar.
This result is invariant to
assumptions
about the mass-luminosity relationship, the mass-loss rates, and rotation
speeds.
In addition, 
from this analysis, we find the age of the Arches cluster to be
2-2.5~\Myr,
assuming coeval formation.

\end {abstract}


\section {Introduction}

The determination of metal abundances as a function of time and Galactic
location
(e.g. disk, bulge and halo)
provides crucial information for understanding the
formation and evolution of our
Galaxy. Such studies  are required to constrain the influence
of the initial mass function, star formation rates,
infall and outflow rates and stellar yields in the galactic models.
The metal content appears 
depleted in the halo and ancient populations of 
globular clusters and displays a roughly solar value in
the metal-rich bulge \citep{fro99,fel00}.
So far,  models
have predicted the presence of a metallicity gradient in the Galactic 
disk with lower metallicity at increasing galactocentric distances.
This result has been confirmed by several studies using different diagnostic
tools. Good agreement has been found among results from HII regions 
\citep{affler97}, planetary nebulae \citep{mac99},
nearby cool stars in the galactic disk \citep{fuhr98} and
early stars (B-type) between 2.5 and 18kpc \citep{roll00,sma01}.

Whether the Milky Way follows the trends observed in other spirals 
\citep{urba04,kenni03}
and reaches
its highest metallicity value in its center
or if on the contrary our Galactic center
is unrelated to the starforming galactic disk is still a source of 
controversy.
Based on measurements of the gas-phase, 
 \citet{shi94} obtained twice solar metallicity 
 from argon and nitrogen emission lines,
 while a solar abundance was derived for neon.
   Further, \citet{mae00} recently obtained four times solar
 abundance by fitting
 the X-ray local emission around Sgr~A~East.
 On the other hand, for the cool stars,
 and based on LTE-differential analysis with other cool supergiants,
 \citet{car00} and \citet{ram97,ram99,ram00} have obtained 
 strong indications for a solar Fe abundance.
It is therefore crucial to obtain metallicity estimates from objects
providing  direct diagnostics, such as
hot stars, and confront them with those 
from the cool-star and gas-phase analyses.
Spectroscopic studies of photospheres and winds of massive hot stars
are ideal tracers of metal abundances because they provide the
most recent information about the natal clouds and
environments where these objects formed.

The Galactic center is clearly a unique environment in the Galaxy. It
hosts three dense and massive star clusters that have recently formed
in the inner 50~pc.
The Arches Cluster \citep{fig99,fig02} is the youngest and densest  cluster
at the Galactic center containing thousands of stars, including at least
160 O stars and around 10 WNLs 
\citep[WN stars still showing H at their surface,][]{chi86}.
The
cluster is very young ($\leq$2.5Myr), and the only emission line stars present
are WNLs and OIf$^+$ having infrared spectra 
dominated by H, \HeI, \HeII, \NIII\ lines. Some have weak  C\,{\sc iii/iv} lines.
The absence of late B-supergiants and LBVs prevents one from
obtaining direct estimates of the important $\alpha$-elements vs. Fe
metallicity ratio
via the \FeII, \SiII\ \&\ \MgII\ lines, as can be done for older clusters like
the Quintuplet \citep{naj04} or the Central parsec cluster.
Being the youngest cluster at the Galactic center, any hint about 
its metallicity would constitute our ``last-minute'' picture of chemical
enrichment of the central region in the Milky Way.

\begin{figure*}
\plotone{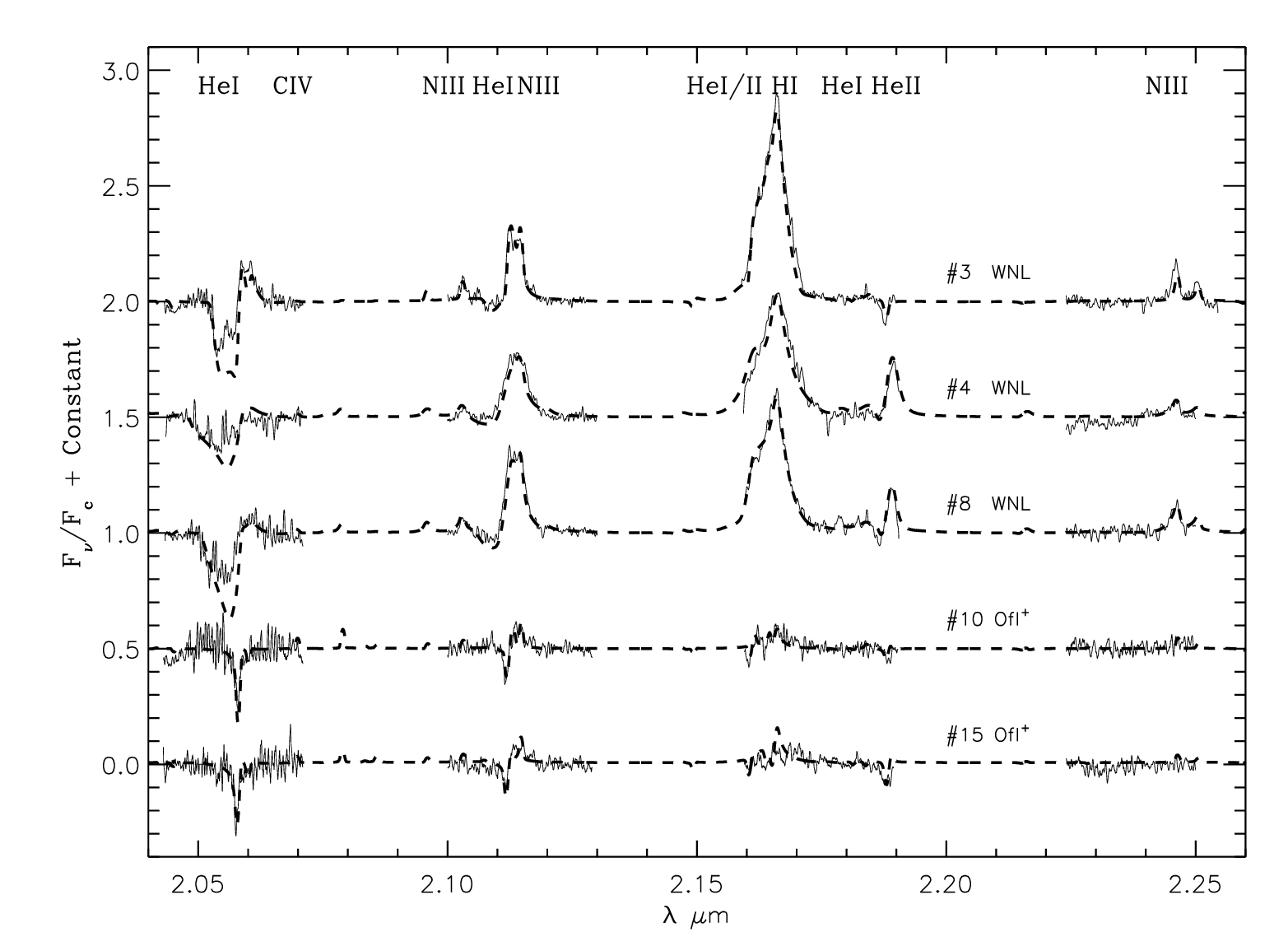}
\caption
{\label{fig:spectra}
Observed spectra ({\it solid}) and model fits ({\it
dashed}) for the three WNL and two OfI$^+$ Arches stars.
The WNL stars generally have strong \ion{He}{1} lines at 2.058~\micron,
2.112~\micron, 2.166~\micron, and 2.189~\micron, and
weaker \ion{N}{3} lines at 2.104~\micron, 2.115~\micron, 2.248~\micron,
and 2.250~\micron. \ion{H}{1} contributes
to the 2.166~\micron\ line. The OfI$^+$ stars have \ion{He}{1} absorption
lines at 2.058~\micron\ and 2.112~\micron,
\ion{N}{3} emission lines at 2.115~\micron, and, perhaps, weak \ion{H}{1}
emission at 2.116~\micron.
Model parameters are listed in Table~1 and objects are given 
according to \citet{fig02}}
\end{figure*}

In this paper, we perform quantitative infrared spectroscopy of the Arches
cluster and provide a robust method to derive metallicity 
in WNL stars using their surface nitrogen abundances and the properties of 
stellar evolution models. 

\section {Observations and Data Reduction}

The observations and data reduction are described in \citet{fig02}, and a brief
summary is provided below.
The spectroscopic observations were obtained on July 4, 1999, using NIRSPEC, the
facility near-infrared spectrograph, on the Keck II
telescope  \citep{mcl98}, in high
resolution mode, covering {\it K}-band wavelengths ($1.98~\micron$ to
$2.28~\micron$). The long slit (24\arcsec) was positioned in a north-south
orientation on the sky, and a slit scan covering a 24\arcsec$\times$14\arcsec\ rectangular region 
centered on the Arches cluster 
was made by offsetting the telescope by a fraction of a slit width to the west between
successive exposures. The resolving power was R$\sim$23,300 (=$\lambda/\Delta\lambda_{\rm FWHM}$), 
using the 3-pixel wide slit, as measured from unresolved arc lamp lines. 

Quintuplet Star \#3  (hereafter ``Q3''), which is featureless in this
spectral region  \citep[Figure~1]{fig98}, was observed as a telluric standard  \citep{mon94}. 
Arc lamps were observed to set the wavelength scale. The spectra were reduced using IDL and IRAF routines. 

The data were bias and background subtracted, flat fielded, corrected for bad pixels, 
and transformed onto a rectified grid of data points spanning linear scales in
the spatial and wavelength directions.

In addition we make use of  HST/NICMOS narrow band photometry 
from \citet{fig02}, which provides equivalent width measurements 
of P$\alpha$.

\section{Models \&\ Discussion}

\begin{figure*}
\plotone{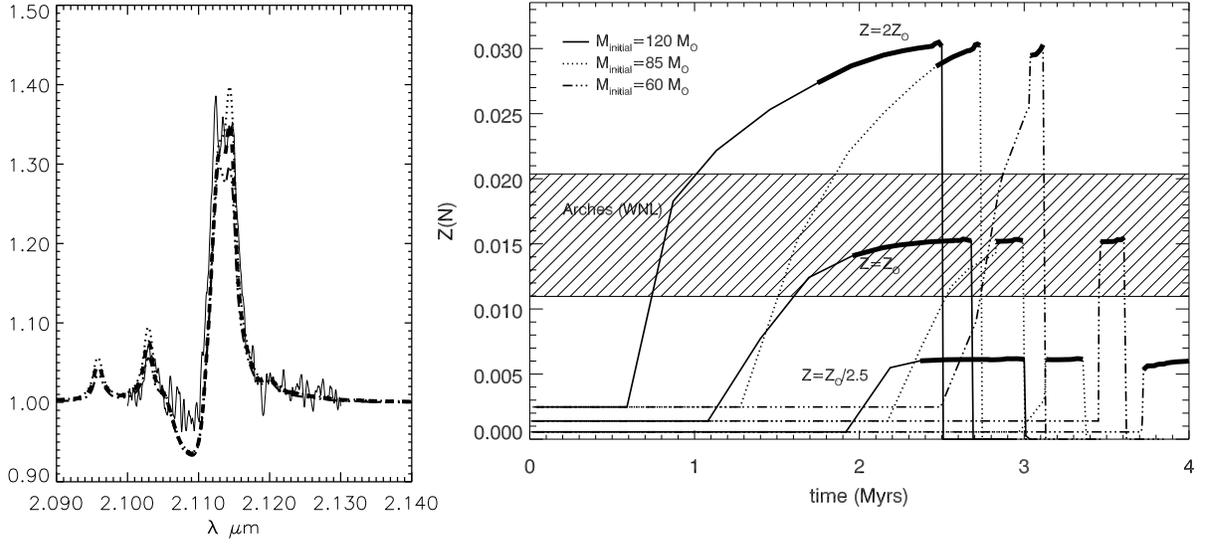}
\caption
{\label{fig:nitrogen}
{\it{Left}}. Leverage of error estimates on N abundance.
Shown is observed 2.10-2.13~\micron\ region (solid)
together with current best fit (dashed) and 30\% enhanced (dotted) and 
30\% depleted (dashed-dotted) nitrogen mass fractions. The \NIII\ 
2.247-2.251~\micron\ doublet (not shown here) displays even higher sensitivity.
{\it{Right}}. Plot of nitrogen mass abundance versus time using
Geneva models \citep{scha92,charb93} for
stars with initial masses of 60, 85, and 120~\Msun, and for initial
metallicities of
\Zsun/2.5, \Zsun, and 2\Zsun; we assume canonical mass-loss rates and no
rotation.
The bold lines correspond to the WNL stellar evolution phase, when the
nitrogen
mass surface abundance reaches its maximum.
The derived nitrogen mass abundances and corresponding errors 
for the Arches WNLs are
cross-hatched.
The measurements require solar metallicity and an age of 2-2.5~\Myr.
}
\end{figure*}

To model the massive Arches stars and estimate their physical parameters,
 we have used the iterative, non-LTE line blanketing method presented
by \citet{hil98}. The code solves the radiative transfer equation in the co-moving frame
 for the expanding atmospheres of early-type stars
in spherical geometry, subject to the constraints of statistical
and radiative equilibrium.
Steady state is assumed, and the density structure is set by the
mass-loss rate and the velocity field via the equation of continuity.
We allow for the presence of clumping via a clumping law characterized
by a volume filling factor f$(r)$.
The  model is then prescribed by the stellar radius, \Rstar,
the stellar luminosity, \Lstar, the mass-loss rate \Mdot,
the velocity field, $v(r)$, the volume filling factor, {\it f}, 
and the abundances of the elements considered.
\citet{hil98,hil99} present a detailed
discussion of the code. For the present analysis we have assumed the atmosphere to be composed of
H, He, C, N, O,  Si and Fe.

 Observational constraints are provided by the K-Band spectra of
 the stars (see Figure~\ref{fig:spectra}) and the narrow band HST/NICMOS
 photometry (filters \fnj, \fnh\ \&\ \fnk) and P$\alpha$ equivalent width (filters \Fnp\ \&\
 \Fnc). Object identifications are 
 given according to \citet{fig02}.

The reduced spectra 
and model fits are shown in Figure~\ref{fig:spectra}. The
top three spectra correspond to some of the most luminous stars in the
cluster. As described in \citet{fig02}, these are nitrogen-rich Wolf-Rayet
stars with thick and fast winds. 
The bottom
two spectra in the figure correspond to slightly less evolved stars with
the characteristic morphology of OIf$^+$ stars \citep[e.g. HD151804,][]{crow97}. 
The main diagnostic H and He lines in the observed K band portions 
(see Figure~\ref{fig:spectra}) are
Br$\gamma$, the \HeI\ lines at  2.058~\micron, 2.060~\micron, 2.112/3~\micron,
2.160-66~\micron\ and
2.181~\micron\ and the \HeII\ lines at 2.165~\micron, 2.189~\micron.
The main carbon lines which may be used to place upper limits
on the carbon abundance are \CIV\ 2.070~\micron\ and \CIII\ 8-7 at 2.114~\micron.
Of concern are the \NIII\ 8-7 lines at 2.103~\micron\ and 2.115~\micron\ as well
as the \NIII\ 5p$\,^{2}$P--5s$\,^{2}$S doublet at 2.247~\micron\ 
and 2.251~\micron.
\citet{fig97} showed that these \NIII\ lines 
appear only for a narrow range of temperatures and wind densities, which occur
in the WN9h (WNL) stage.
The fairly distinct nature and energies of the multiplets
involved in each of both \NIII\ line sets provide strong constraints
for the determination of the nitrogen abundance. Thus, at the S/N of our
spectra, our models show that the WNLs \NIII\ lines can easily track relative
changes as low as 20\% in the nitrogen abundance, and a 30\% error should
be regarded as a safe estimate, as shown in 
Figure~\ref{fig:nitrogen}-{\it{left}}.

Individual stellar parameters of the WNL stars
(objects \#3,\#4 \&\ \#8)
 are listed in Table~1, together with those derived 
 for two OIf$^+$ stars (\#10 \&\ \#15) in the
 cluster. It is important to note that given the extreme sensitivity 
 of the \HeI\ and \HeII\ line
 profiles in this parameter domain to changes in effective temperature, 
 we estimate our errors to be below 1000~K. For the WNLs, 
 the clumping factor is
 basically determined by the red emission wing of  Br$\gamma$, while
 \Vinf\ is set by the \HeI~2.058~\micron\ line.
 For the 
 OIf$^+$ stars,   
 we investigated
 the effects of varying both the clumping factor and the terminal velocity.
 Only an upper limit is obtained for the latter from the 
 shape of the Br$\gamma$-He complex at 2.16~\micron, while there was no
 unique solution for {\it f}. Therefore we adopted f=0.1, which is 
 roughly the average value found for the WNLs.
 For all objects, the relative strength
 between the H and \HeI\ lines is used to derive the He/H ratio while 
 their absolute strengths provide \Mdot.
 The stellar parameters displayed in
 Table~1 (\Mdot, He and N abundance) reflect the striking difference in
 morphology between the observed WNL and OIf$^+$ spectra.
Of particular importance  is the roughly same surface abundance fraction of N, Z(N), obtained
for all three WNL objects ($\sim$1.6\%) well above the upper limit found
for the 
OIf$^+$ stars ($\sim$0.6\%).
WNL stars do not exhibit any primary diagnostic line in their
K-Band spectra to 
estimate metallicity. However, the crucial r\^ole of Z(N) 
in determining metallicity from WNL stars can be immediately anticipated if
we make use of the stellar evolution models for massive stars.

According to the evolutionary models by \citet{scha92} and \citet{charb93},
a star entering the WNL phase still shows H at its surface
together with strong enhancement
of helium and nitrogen and strong depletion of carbon and oxygen as expected
from processed CNO material. Further,
once a massive star reaches this phase,
it maintains a nearly constant
Z(N) value throughout this period, and the amount achieved essentially depends
only linearly on the original
metallicity (see Figure~\ref{fig:nitrogen}-{\it{right}}). Namely, the
maximum Z(N) in the WNL phase is set
by the initial content of CNO (sum of C, N and O mass fractions)
present in the natal cloud.
However, since we expect the CNO abundance
in the natal cloud to
scale as the rest of metals, the nitrogen 
surface abundance must trace the metallicity of the cluster.
Further, this value is 
basically unaffected by the mass-loss rate assumed and the presence of 
stellar rotation during evolution \citep{mey04}.
Hence, once we clearly identify a star to be on the WNL evolutionary phase,
we may confidently use its N surface abundance fraction to estimate
metallicity. The He/H ratio, the upper limit on the very depleted
carbon abundance and the \Teff\ and \Mdot\ values indicate that this is
the case for objects \#3, \#4 \&\ \#8. The derived
Z(N) ($\sim$1.6\%) is the one expected for 
{\bf solar} metallicity from the evolutionary models. Due to the linear
behavior of  Z(N) with metallicity, we make use of the uncertainties
in the N abundance estimate to obtain an error of roughly 30\% for the
metallicity (see Figure~\ref{fig:nitrogen}-{\it{left}}).
The reliability of our method is  demonstrated in
Figure~\ref{fig:nitrogen}-{\it{right}},
where we display the nitrogen mass fraction as a function
of time for stars with initial masses of 60, 85, and 120~\Msun,
and metallicities
equivalent to 2, 1, and 0.4 times solar, assuming the canonical mass-loss
rates \citep{scha92}. The cross-hatched region represents our estimates including
the above quoted errors of the
nitrogen mass fraction for the Arches WNLs.
To test the validity of the nitrogen abundance method to trace metalliticy,
we may compare our results with previous abundance estimates for these
objects in regions of different metallicity. Indeed, \citet{crow00} found
that the nitrogen surface abundance value reached during the WNL phase 
in the SMC (0.3\%) reproduced very well the observed spectra of SK-41 while,
\citet{crow95} found surface abundances corresponding to solar metallicity
in their analysis of galactic WNL stars.

 Another important direct result from Figure~\ref{fig:nitrogen} and 
 Table~1 is a tight estimate for the masses and ages of the objects.
 The luminosities of \#4\ \&\ \#8\ are consistent with initial masses of
 $\sim$120~\Msun\ for these WNL stars while the luminosity of
 \#15\ seems to be consistent with masses 70-90\Msun\ for the O stars.
 Assuming these masses, equal initial abundances, and coeval formation,
 one finds that both the WNL and O stars require solar metallicity and an
 age of 2-2.5~\Myr. Current luminosities of \#3\ and \#10\ can be
 interpreted in terms of the effects of rotation on the evolution of 
 massive stars. New evolutionary models for massive stars including
 rotation \citep{mey04} show that the WNL phase is reached much faster
 for those objects with high initial rotational velocities.
 Once the star reaches 
 the WNL phase, its luminosity starts to decrease. Thus, for a given
 initial ZAMS mass, a scatter in 
 the initial rotational velocity will result in a range of current
 luminosity.
 
 If our assumption of coeval star formation is correct, the OIf$^+$ star
 \#10\ with current high luminosity, but which is less evolved
 than the WNLs,
 should have had a much lower initial velocity than the rest
 of the WNLs \citep[see tracks by][]{mey04}. While this assumption
 can not be proved, as we have no way to estimate the current rotational
 velocity of the WNLs, a 
 high present rotational velocity for the OIf star \#10\ would definitely
 rule out 
 such a scenario. We use our models to find a consistent maximum 
 rotational velocity of v$sini$$<75km/s$ for \#10\ from
 the narrow \HeI~2.058\micron\ absorption line (see Fig.~\ref{fig:spectra}).  
 Likewise, 
 the low current luminosity of the WNL star \#3\ would be consistent with a
 slightly lower 
 ZAMS mass than \#4 and \#8 and a higher initial rotational velocity
 \citep[as shown in the evolutionary tracks by][]{mey04}.

 Our result of solar metallicity
 runs counter to the trend in the disk \citep{roll00,sma01} 
 but is consistent with the findings from cool star studies
 \citep{car00,ram97,ram99,ram00}. This may imply that the ISM in
 the disk does not extend inward to the GC, or that 
 the GC stars are forming out of
 an ISM that has an enrichment history that is distinctly different from that
 in the disk. Our result is more consistent with the values found for
 the bulge \citep{fro99,fel00}.
 Further results for other GC regions, like the Central cluster or the
 Quintuplet cluster \citep{naj04} are needed.
 In a forthcoming paper we will investigate the wind properties of a large
 sample ($\sim$20 objects) of WNL and OIf stars in the Arches cluster.

\acknowledgements
We thank George Meynet and Andre Maeder for useful discussions.
F. N. acknowledges grants AYA2003-02785-E and ESP2002-01627.


\newpage

\begin{deluxetable}{llllll}
\tablecaption{WNL and O stars in the Arches Cluster}
\tablehead{
\colhead{{Parameter}} &
\colhead{{\#3}} &
\colhead{{\#4}} &
\colhead{{\#8}}  & 
\colhead{{\#10}} &  
\colhead{{\#15}}} 
\startdata
\Rstar (\Rsun)           & 43.5 &39    & 48    &  48    & 29    \\
\Lstar ($10^{5}\Lsun$)   &10.3  &16.5  & 18.5  &  18.5  & 5.85  \\
\Teff  ($10^{4}$K)       &2.79  &3.32  & 3.09  & 3.07   & 2.95  \\
He/H                     & 0.50 & 0.57 & 0.67  & 0.33   & 0.33   \\
Z(He)                    & 0.65 & 0.68 & 0.71  & 0.56   & 0.56   \\
Z(N)                     & 1.7  &1.4   & 1.6   & 0.6:   & 0.6:  \\
Z(C)                     &0.02: &0.03: & 0.02: & 0.08:  & 0.15: \\
\Mdot ($10^{-6}\Msunyr$) &21.5  &32.5  & 45.0  & 4.3    & 9.11  \\
\Vinf (\kms)             &840   &1400  & 1100  & 1000:  & 1000: \\
Clumping factor {{f}}    &0.1   & 0.15 &0.08   & 0.1    & 0.1   \\
$\eta$=\Mdot \Vinf / ( \Lstar /c )  &0.72  &1.35  & 1.32  & 0.11   & 0.75  \\
LogQ$_{H^{+}}$           & 49.4 &49.9  & 49.9  & 49.8   & 49.2  \\
LogQ$_{He^{+}}$          & 47.3 &48.6  & 48.5  & 48.4   & 47.3  \\
\enddata
\tablecomments{
Object identifiers after \citet{fig02}.
Upper limits are quoted as ({\bf{:}}).
He/H is ratio by number and other abundances are mass fractions.
$\eta$ is the performance number and Q  is the ionizing photons rate in photons/s.}
\end{deluxetable}

\end{document}